# Dynamics-based peptide-MHC binding optimization by a convolutional variational autoencoder: a use-case model for CASTELO

David R. Bell[1+], Giacomo Domeniconi[1,2+], Chih-Chieh Yang[1], Ruhong Zhou[1], Leili Zhang[1*], and Guojing Cong[1*]

1. IBM Thomas J. Watson Research Center, Yorktown Heights, NY 10598, USA
2. IBM Research, Saeumerstrasse 4, 8803 Rueschlikon, Switzerland

+ Contributed equally to this work

* To whom correspondence should be addressed. E-mail: zhangle@us.ibm.com (L.Z.), gcong@us.ibm.com (G.C.)

**Abstract**

An unsolved challenge in the development of antigen specific immunotherapies is determining the optimal antigens to target. Comprehension of antigen-MHC binding is paramount towards achieving this goal. Here, we present CASTELO, a combined machine learning-molecular dynamics (ML-MD) approach to design novel antigens of increased MHC binding affinity for a Type 1 diabetes-implicated system. We build upon a small molecule lead optimization algorithm by training a convolutional variational autoencoder (CVAE) on MD trajectories of 48 different systems across 4 antigens and 4 HLA serotypes. We develop several new machine learning metrics including a structure-based anchor residue classification model as well as cluster comparison scores. ML-MD predictions agree well with experimental binding results and free energy perturbation-predicted binding affinities. Moreover, ML-MD metrics are independent of traditional MD stability metrics such as contact area and RMSF, which do not reflect binding affinity data. Our work supports the role of structure-based deep learning techniques in antigen specific immunotherapy design.

**Introduction**

Antigen specific immunotherapies for autoimmune diseases act through several mechanisms including inducing T-cell deletion/tolerance, eliciting anti-inflammatory T-helper 2 cells[1], and blocking the HLA binding groove[2-3], among others[4]. Peptide-MHC-TCR (pMHC-TCR) binding interactions are complex[5-8], but comprehension of peptide-MHC binding is of paramount importance for therapeutic design. There are numerous *in silico* approaches to predict pMHC binding, with the majority combining sequence-based informatics and machine learning techniques to predict binding for select HLA serotypes[9-12]. Although the reported accuracy of these techniques for pMHC binding prediction is quite high[13-14], confidence in MHC-II binding predictions remains lower, partially due to scarcity of data, ambiguity of binding registers and variable contributions of flanking regions[9, 15-16]. Structure based modeling of pMHC binding is able to overcome these limitations by incorporating three-dimensional atomistic representations rather than linear sequences[17-19]. Docking methods employing charge interactions and shape complementarity have found success in pMHC-TCR binding studies[20-22], but are unable to account for dynamics. Molecular Dynamics (MD) simulations can explicitly account for important pMHC dynamic behavior such as register shifts, solvent dynamics, and HLA residues changing orientations within the binding cleft[8, 23-25]. MD simulations have been used previously to depict pMHC-TCR interactions[26-30] and remain a rigorous standard for modeling pMHC binding.

The analysis of MD simulations is often constrained to manual identification of solute interactions by domain experts, posing an opportunity for Machine Learning (ML) techniques to characterize chemical interactions in an automated and less-biased manner. Autoencoders, in particular, have provided advanced insights into MD trajectories[31-34]. Autoencoders are unsupervised neural networks widely used for dimensionality reduction and pattern recognition. Their architecture is composed of an encoder part that learns and represents an input into a latent space and a decoder part that reconstructs the original input using the low-dimensional features. Variational autoencoders (VAEs) introduce the optimization constraint to the latent space to be normally distributed, this coerces the network to more evenly distribute the information into the latent space[35]. For MD simulation analysis, a common VAE input is a contact matrix from a single MD time step; hence, convolutional layers (CVAE) rather than regular feedforward are typically used. This leads to filter maps that can better recognize local patterns independently of the position.

In this work, we expand CASTELO[34] by combining a CVAE ML pipeline with MD simulation to predict antigen residue binding contributions and optimize antigen-HLA binding for antigen specific immunotherapy design. We have previously shown using the CASTELO method that autoencoders trained on MD trajectories can identify unstable subdomains in small-molecule drug candidates for lead optimization[34]. In this study, we expand this methodology by incorporating deep learning methods to protein-protein interactions of the pMHC complex. We target a previously reported[36] Type 1 Diabetes (T1D)-implicated cell population that expresses both BCR and TCR. We train our model on 48 systems consisting of 4 HLA serotypes, 4 different antigens, and 3 binding registers per antigen. Following 500ns MD simulation for each system, we apply a CVAE pipeline to predict antigen residue contributions to binding. Based on these results, we optimize antigen residues by computing mutational binding affinities using Free Energy Perturbation (FEP) calculations. We report several gain-in-affinity residue mutations for antigen specific immunotherapy design as well as novel structure-based machine learning metrics for pMHC binding.

## Results

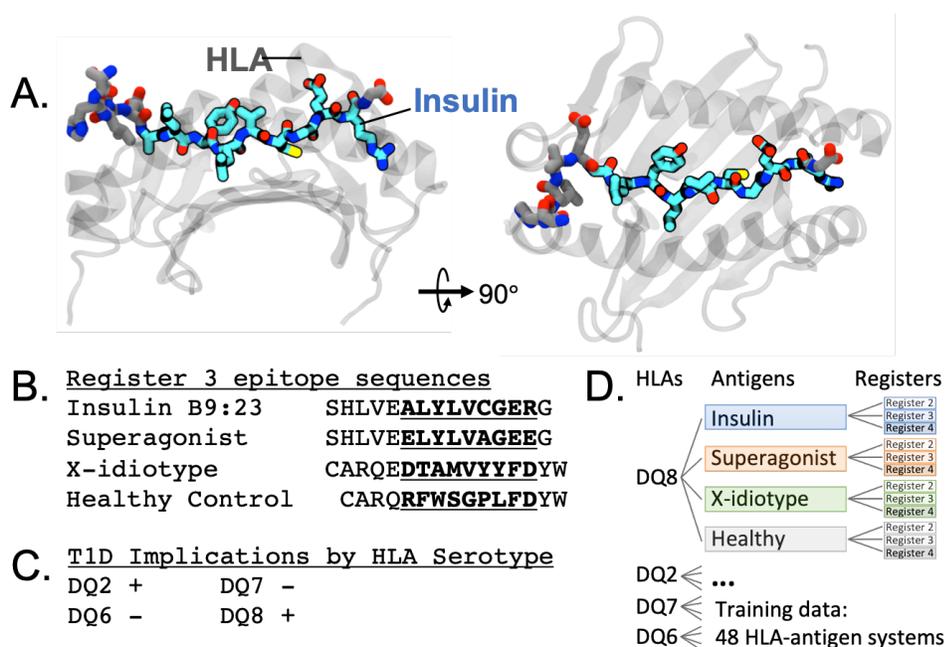

Figure 1. HLA-antigen systems studied. (A) HLA-antigen representative structure illustrated using HLA-DQ8 and Insulin B9:23 peptide in register 3. The core epitope is colored cyan while the flanking domains are in gray. Note that only the HLA-α1 and -β1 subunits are shown for clarity; HLA-α2 and –β2 subunits were included in all MD simulations. **(B)** Studied antigen sequences with register 3 core epitope in bold underline. **(C)** Studied HLA serotypes marked according to their T1D disease propensity, + indicates increased risk, - indicates reduced risk[37-38]. **(D)** HLA-antigen and register system combinations. With 4 HLA serotypes, 4 antigens and 3 registers, a total of 48 HLA-antigen systems are included in training data.

Figure 1 presents the MHC-II-antigen systems studied. Four antigens were investigated: the canonical insulin B:9-23 peptide, a superagonist developed by Dai and coworkers[39], and two sequences taken from ref[36]: the X-idiotype and a healthy control sequence not observed in T1D patients. Each antigen was simulated in three registers, labeled 2-4, with register 3 corresponding to the TCR-recognized insulin B:9-23 register[40] and register 2 being one residue shift to the left while register 4 is one residue shift to the right. Antigen binding was investigated for four HLA serotypes: DQ2, DQ6, DQ7, and DQ8 with DQ2 and DQ8 holding enhanced risk for T1D and DQ6 and DQ7 holding reduced risk for T1D[37-38]. In total, 48 systems (4 HLAs x 4 antigens x 3 registers) were modeled using Molecular Dynamics (MD) simulations, each for 500ns for a total of 24μs (see Fig. S1 for representative RMSD profiles). Note that Figure 1 only shows the HLA-α1 and -β1 subunits, while the full HLA (α1, α2, β1, β2) molecules were modeled in MD simulations. Three simulation systems (X-idiotype, healthy control, and superagonist with HLA-DQ8) were previously reported in ref[36].

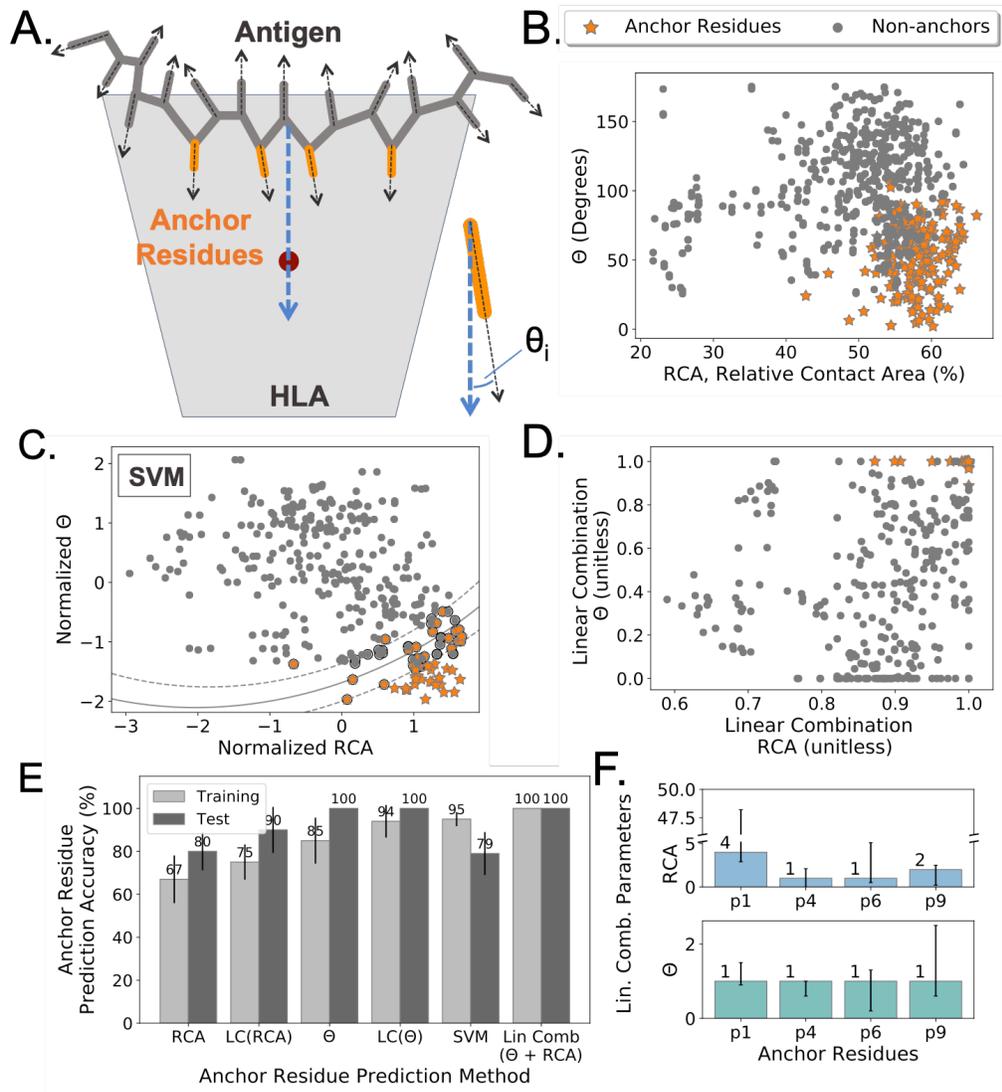

**Figure 2.** Anchor residue determination. **(A)** HLA-antigen cartoon illustrates anchor residues in orange and metric θ. $\theta_i$ is defined as the angle between $i^{th}$ residue vector and a vector from the antigen center to the HLA center. **(B)** Distribution of anchor residues and non-anchor residues as a function of $\theta_i$ and HLA-residue contact area. HLA-residue relative contact area (RCA) is normalized to the total surface area of residue i. **(C)** SVM decision boundary (solid line) and margins (dashed lines) to classify anchor vs. non-anchor residues. SVM mean accuracy is 79%/95% test set/training set. **(D)** Plot showing linear combination of residue contact areas and $\theta_i$ which resulted in 100% accuracy of anchor residue classification for the systems studied. **(E)** Accuracy of anchor residue prediction methods. SVM classifier used both θ and RCA. LC stands for linear combination. Highest accuracy in both training and test sets was a linear combination of θ and RCA. Error bars represent 95% CI. **(F)** Linear combination coefficients of θ and RCA for 100% anchor residue prediction accuracy. Error bars indicate possible range of coefficients to still maintain 100% prediction accuracy. Sample sizes: 48 training, 20 test.

Our novel method of anchor residue identification is presented in Figure 2. Anchor residues are antigen residues whose side chains are buried inside deep pockets in the MHC binding cleft. Due to their large interaction with the HLA, anchor residues make excellent mutation candidates for binding affinity enhancement[41] or destruction[42-43] and thereby immunotherapy optimization[41-43]. Anchor residues are readily identified from visual inspection of experimental structures using molecular visualization programs[44-45]. However, for large, structure-based pMHC binding data sets, there is a need for automated identification of anchor residues, which is what we address here, focusing on MHC-II binding. MHC-II antigens range in length from 13-25 residues[46], with a 9-residue core epitope that binds inside the HLA binding cleft. Of this 9-residue epitope, also called a register, residues 1,4,6, and 9 are typically anchor residues. We sought to identify anchor residues and thus the epitope register based on structural metrics of the MHC-II-antigen system.

In order to identify anchor residues from a novel HLA-antigen structure, we combined antigen residue-HLA orientation, $\theta$, with antigen residue-HLA contact area. The definition of $\theta$ is shown in Figure 2: for each antigen residue, a side chain vector was defined from the C$\alpha$ atom to the terminal side chain heavy atom (Hydrogen for Glycine). Another vector is defined from the antigen center to the HLA center. $\theta$ is then defined as the angle between the side chain vector and the antigen-HLA vector. Low $\theta$ values indicate that the residue side chain is pointed toward the HLA like an anchor residue, while higher $\theta$ values indicate that the residue side chain is pointed away from the HLA. Similarly, anchor residues should have high relative contact area (RCA). However, as shown in Fig. 2B, anchor residues cannot be naively identified using these two $\theta$ and RCA metrics.

As a reference point to identify anchor residues vs. non-anchors, we built a Support Vector Machine (SVM) model, shown in Fig. 2C. This classifier identified anchor residues with about 95% mean accuracy for the training set. We then built a test set of 20 MHC-II-antigen experimental structures and compared the accuracy of the SVM to the $\theta$ and RCA metrics, as well as empirically-determined linear combinations of $\theta$ and RCA, Fig. 2E. The details of the test set are shown in Table. S1. As seen in Fig. 2E, each anchor residue identification method achieved high accuracy in the test set, except for the SVM. Residue orientation $\theta$ alone was sufficient to predict anchor residues in the test set. However, the highest accuracy in both test and training sets came from a linear combination of residue orientation $\theta$ and contact area RCA. The coefficients of the linear combination are shown in Fig. 2F. The residue orientation $\theta$ remains unmodified with coefficients of 1, while the first and last anchor residues have increased contact area RCA coefficients than the middle two anchor residues. Interestingly, the first anchor residue RCA coefficient can be as high as 48 (Fig. 2F, error bar) and still result in 100% anchor residue prediction accuracy for the linear combination method. One limitation of our anchor residue classification methods is the small training and test set sizes (48 and 20 respectively). This indicates that the classification methods presented here are likely overfit to the systems of study and should be re-evaluated with additional availability of experimental structures. Nevertheless, these results indicate several potential methods of naïve anchor residue identification for ML immunotherapy workflows.

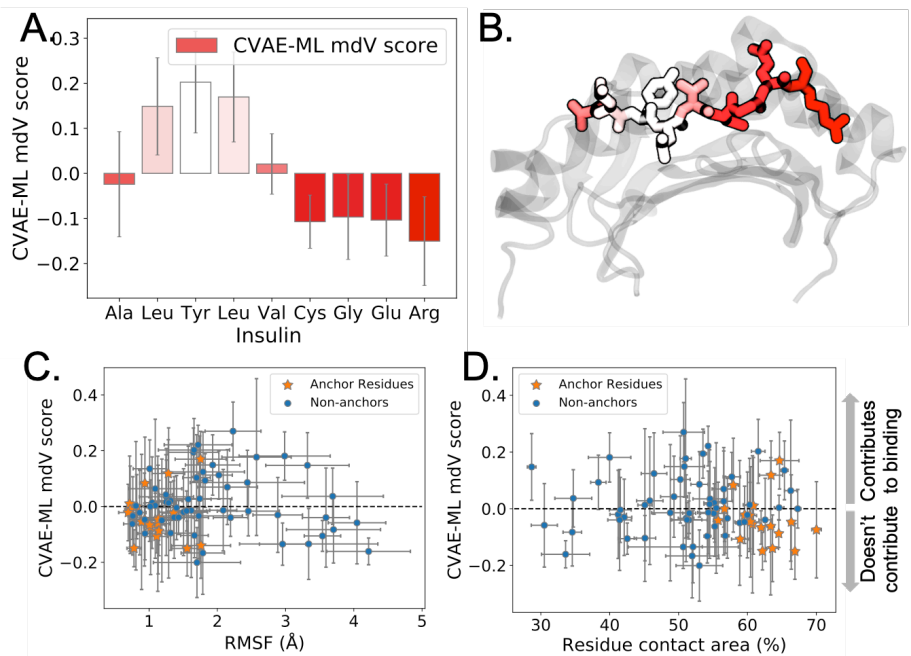

**Figure 3.** Representative CVAE analysis of epitope residue stability. **(A)** CVAE-ML *mdV* score for the HLA-DQ8-insulin antigen register 3 system. Red coloring reflects contribution to epitope binding, red: less contribution, white: more contribution. **(B)** CVAE-ML *mdV* score mapped onto the HLA-DQ8-insulin antigen register 3 structure, showing instability in the antigen C-terminus (red). Same coloring as **(A)**. **(C)** CVAE-ML *mdV* score as a function of residue fluctuations, RMSF. **(D)** CVAE-ML *mdV* score as a function of residue contact area. The lack of correlation in **(C-D)** indicates the CVAE is extracting structural information distinct from residue fluctuation and contact area. All error bars represent 95% confidence intervals.

MD simulations typically generate extensive amounts of high-dimensional data that are difficult to analyze. Many ML techniques have been used to build low-dimensional representations of MD simulations. In our approach, we propose the use of autoencoders (AE), a well-known neural network architecture for dimensionality reduction tasks without the need of hand-crafted feature engineering. In particular, our goal is to expand the CASTELO method by using AEs in combination with unsupervised clustering methods to identify antigen residue binding contributions from a large, high-dimensional MD dataset of pMHC simulations.

CVAE-ML results for the insulin B:9-23 epitope binding HLA-DQ8 in register 3 (SHLVE**ALYLVCGER**G) are presented in Fig. 3. For simplicity, clustering results of the CVAE latent space were distilled into a singular residue metric, termed the *dV* score (detailed in the Methods). For simplicity, we present a modified *dV* score (as described in the Methods), *mdV*, to predict the residue contribution to binding proportionally from positive to negative values. A positive *mdV* score indicates the residue contributes to the binding mode, while a *mdV* score of 0 indicates average contribution, and a negative *mdV* score indicates the residue does not contribute to the binding mode. The magnitude of the *mdV* score matters: a basic assumption of our work is that lower *mdV* scoring residues can be mutated for greater binding affinity while higher *mdV* scoring residues cannot be mutated for improvement. As shown in Fig. 3A-B, residue p9 Arg of the insulin epitope has a low *mdV* score and is hence predicted to not contribute to binding. This is supported by experiment which shows that a positively charged residue at site p9 clashes with charged HLA residues, and that mutation to a negatively charged residue improves binding interaction[39]. As seen in Fig. 3C-D, the CVAE-ML *mdV* score does not correlate to either the residue root mean square fluctuation (RMSF) or residue contact area. This indicates that the *mdV* score captures behavior independent of the RMSF and the residue contact area. RMSF and residue contact areas are commonly used to assess residue stability; however, as shown in Fig. S2, RMSF and contact area serve as poor predictors for residue binding contributions. The CVAE-ML *mdV* score predicts residue contribution to binding mode separate to stability metrics and thus represents a new technique for binding mode prediction.

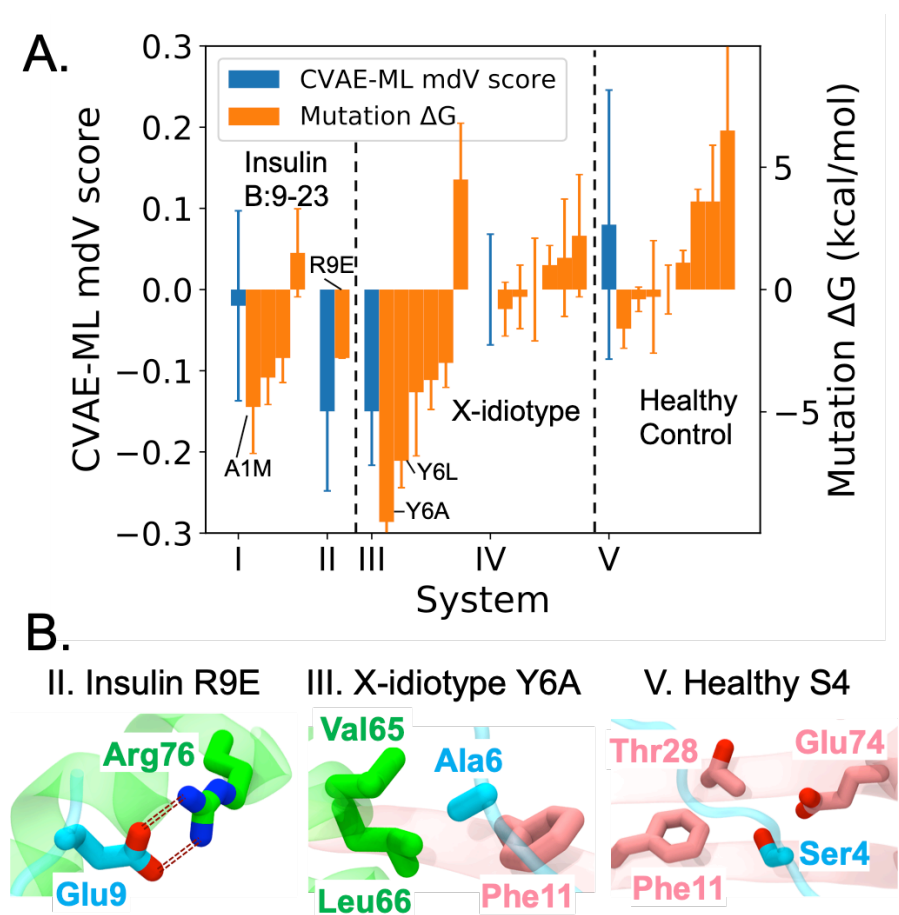

**Figure 4.** CASTELO results for novel antigen design. **(A)** CVAE-ML *mdV* score (blue) in comparison to mutational binding affinity changes (orange). System I is insulin p1A, II is insulin p9R, III is X-idiotype p6Y, IV is X-idiotype p7Y, and V is healthy control p4S, all systems are with HLA-DQ8 in register 3. All mutational binding affinities (orange bars) are computed from free energy perturbation calculations, except for system II taken from literature[39-40]. Labels include antigen identities (separated by dashed lines) and favorable mutations. Error bars represent 95% confidence intervals. **(B)** Structures of optimized favorable mutants for systems II and III and the non-improvable system V.

Fig. 4 presents CASTELO-predicted novel antigen design results. Five residue systems were selected for CVAE-predicted optimization: I. p1 Ala and II. p9 Arg from HLA-DQ8 insulin register 3, III. p6 Tyr, and IV. p7 Tyr from HLA-DQ8 X-idiotype register 3, and a positive control V. p4 Ser from HLA-DQ8 healthy epitope register 3. Optimization of these residues was explored through mutational Free Energy Perturbation (FEP) calculations as described in the methods. Note that due to the expense of FEP calculations, a selected subset of residue mutations reflecting a range in residue hydrophobicities and sizes was initially tested and then expanded upon.

Fig. 4B illustrates selected mutations for systems II, III, and V. Mutations for system II. insulin p9 Arg and III. X-idiotype p6 Tyr both act to increase HLA-DQ8 binding affinity. As a reminder, low *mdV* scores indicate the residues do not contribute to binding and can be mutated for enhanced HLA-antigen binding affinity. Both systems are anchor residues with the lowest *mdV* score of their respective epitopes and thus made for promising mutation targets (see Tables S2-S4 for mdV scores of all antigen residues). The insulin R9E mutation has been experimentally verified to result in 100x greater binding interaction[40], nicely agreeing with our *mdV* score results. Likewise, FEP calculations revealed the X-idiotype p6 Tyr residue was readily improved by mutation to smaller residues, with the Y6A mutation holding the largest affinity improvement of $\Delta G = -9.5 \pm 1.3$ kcal/mol ($\pm$ 95%CI). However, mutation to a larger residue, Trp, resulted in worse binding affinity, $\Delta G = 4.5 \pm 2.3$ kcal/mol, shown as the positive orange bar of system III. in Fig. 4A. This illustrates the notion that although the CVAE-ML *mdV* score can predict a residue is not contributing to the binding mode, it does not mean that every mutation will be favorable. Rather, when the CASTELO CVAE-ML model predicts a residue is not contributing to the binding mode, that means there is a *potential* for binding affinity optimization by mutation. The CVAE-ML model, under its current form, does not predict which mutant residue will increase binding affinity. However, the ability to predict favorable mutations is greatly desired by the field and remains an active area of research.

Systems I and IV are predicted by CASTELO to be of roughly average contribution to the binding mode. Mutation free energy calculations for system IV. X-idiotype p7 Tyr do not clearly identify any enhanced binding affinity mutations. Most mutations for p7 remain close to zero or have decreased binding affinity. Free energy calculations and experimental evidence[39] for system I. insulin p1 Ala do reveal favorable binding affinity mutations, even though the CVAE-ML *mdV* score is of roughly average value. When the p1 Ala *mdV* score is compared across the epitope, it is much higher than the last 4 residues of the epitope which have low *mdV* scores (Fig. 3A, Table S3). Locally, however, the p1 Ala *mdV* score is the lowest within the first 5 residues of the epitope and is the main residue to target for N-terminal epitope binding optimization.

System V. healthy control p4 Ser was selected as a positive control that could not be further optimized for binding affinity enhancement. This was well supported by our FEP calculations, which found most mutations decreased binding affinity or remained neutral. A cysteine mutation slightly improved binding affinity: S4C, $\Delta G = -1.6 \pm 0.8$ kcal/mol ($\pm$ 95%CI). However, substantial binding affinity improvements were not observed, supporting the mdV score conclusion that p4 Ser contributes to binding stability and is not a strong candidate for mutation.

## Discussion

Exploration of MHC-II-antigen binding remains an active area of research for immunology and immunotherapy design[5]. Exhaustive sampling of all possible epitopes that could bind MHC-II remains beyond current experimental and computational means. Therefore, directed study and mutation for binding affinity optimization or destruction is highly desired. Current antigen design methods rely on sequence-based ML models[13-14] in combination with domain expertise of researchers[39, 47]. There is a need for automated, structure-based MHC-II-antigen binding exploration and design, which is what we have targeted here. Similarly, Ochoa et al.[24] used structure-based approaches to predict antigen-MHC-II binding affinity by using MD simulations to sample conformations and docking scoring functions to rank binding affinity. This represents a valid approach to compare two or more antigens based on MD simulations, but lacks a straightforward way of identifying which residues are unstable and can be mutated for greater affinity. In comparison, we sought to develop a tool to help design antigen specific immunotherapies and optimize antigen binding by identifying per-residue binding contributions.

The existence of fixed-interval anchor residues whose side chains bury into deep HLA binding pockets is a unique feature of pMHC structure and is exploited by many pMHC binding algorithms[13, 29]. Anchor residues can be rapidly identified from visual inspection of pMHC structures. However, as more pMHC-II experimentally-resolved structures become available, automated determination of anchor residues will be needed for structure-based pMHC-II binding models. We found that a linear combination of antigen sidechain-global HLA orientation combined with relative contact area for each residue was sufficient to identify 100% of anchor residues for both the training and test sets used here. However, this model may suffer from overfitting and will require additional testing as more experimental structures are determined.

Our dynamics-based CVAE ML pipeline CASTELO greatly reduces the time and computational expense required for the design of antigen specific immunotherapies such as HLA-blockers or novel antigen vaccines. In our previous works of antigen mutation and binding affinity improvement[26, 28, 48], typically only 1 in 60-100 mutations is found to improve binding affinity (≈-1.5kcal/mol), with a high cost of computing resources. The CVAE ML pipeline used in this work greatly reduces this expense by focusing our efforts on residues that have the potential to improve binding affinity upon mutation. In this work, for instance, our success rate was around 1 in 2-10 mutations found to be favorable according to FEP calculations.

Lastly, one limitation of this work is that although CASTELO predicts residue contributions to binding, it does not predict what residue mutations would be favorable (alanine or tryptophan, for example). Predicting which residue to mutate to *de novo* would greatly reduce computational modeling expense and remains a topic of ongoing research. Additionally, we compute mutational binding affinities using Free Energy Perturbation (FEP) calculations, which do not account for any structural changes upon mutation. Antigens are intrinsically disordered, capable of adopting multiple conformations even when bound to the MHC[8, 29, 49]. Ideally, we would follow-up each mutation with MD simulation[50]. However, given our experience with FEP for MHC and other systems[36, 51], consistent FEP calculations of point mutations with only moderate error as presented here, typically do not result in drastic structural rearrangements (like 'switch' proteins[52-53]) and agree well with experimental values (within 1kcal/mol[51]).

## Conclusion

In this work, we extend the deep learning CVAE protocol CASTELO to predict per-residue antigen-MHC-II binding contributions and then mutate poorly binding residues to optimize antigen-HLA binding. We also present a structure-based anchor residue classification scheme to determine the binding register which is ambiguous for MHC-II antigens greater than 9 residues. From CVAE results, we find several residues that can be mutated for improved binding affinity as confirmed by experimental literature and novel Free Energy Perturbation (FEP) calculations. Importantly, CVAE results are found to be independent of traditional MD stability metrics such as RMSF and contact area, supporting its role as a novel technique for MD analysis. Our work presents a powerful protocol for antigen-MHC-II binding prediction and design of antigen specific immunotherapies.

## Methods

**MD Simulations.** HLA-antigen systems were initially modeled using MD simulations. As described in our previous work[36], the X-idiotype, mimotope, and healthy control systems were built from structure PDB ID: 5UJT, while the insulin system was taken from PDB ID: 1JK8. The X-idiotype and healthy control antigens were mutated from the 5UJT mimotope antigen using VMD[45]. Each protein system was solvated in TIP3P water molecules with 100mM NaCl salt concentration. CHARMM36[54] force field parameters were used for protein modeling. The systems were minimized for 50,000 steps with protein heavy atom constraints followed by an additional 50,000 steps without constraints. The systems were then equilibrated for 1ns and modeled for 500 ns MD simulations at 310 K with a 2 fs time step. All simulations were conducted using NAMD2[55-56] on an IBM Power8/9 cluster.

    **CASTELO** Convolutional Variational Autoencoders (CVAE) were used to model each MD trajectory in a low-dimensional space. For the ML analysis, MD trajectories consisted of 2500 frames saved every 200ps, for a total of 500ns simulation. Similar to previous works[31-32], we model the input as a contact matrix/dynamism tensor between antigen and protein residues in each structure of the simulation. Specifically, the contact matrix has size of NxM, where N is the number of antigen residues and M is the number of protein residues. Considering our aim of finding binding patterns during the simulation, the temporal information over time is important to locate stable and unstable states. We enriched each time step *t* input using the difference of contacts between the current time *t* and a previous one *t-delta*. In our pipeline each time step of the simulation is then modeled as a tensor of 2xNxM, with the first two dimensions representing the contact matrix in *t* and the difference of the contacts in *t* and *t-delta*. We used CVAE to represent each contact tensor in a d-dimensional latent space, i.e. a vector of d elements. This intrinsically brings time steps with similar contact matrices to closer positions in the latent space, allowing clustering techniques to group sets of time steps with similar behavioral patterns. In other words, applying clustering over the latent space vectors generated

by CVAE allows the discrimination of big clusters with several time steps of similar behavior, i.e. stable situation, or small clusters, i.e. unstable. We used HDBSCAN[57] as our clustering methodology. This algorithm overcomes the limitation of knowing in advance the number of clusters, that is the typical drawback of partitioning techniques such as K-means, and also the density threshold, typically required by the standard DBSCAN.

Our goal is to design novel antigens by proposing favorable mutations of the antigen residues; for this purpose, we used a comparison between the overall antigen binding behavior and that of each individual residue. In our pipeline, we firstly used CVAE and HDBSCAN using the 2xNxM contact tensor to create a reference clustering result, i.e. an overall metric of the stable states during the simulation. Then, we applied the same analysis but only focusing on the contacts of each individual residue, i.e. using a contact matrix of size 2x1xM for each time step. We repeated this procedure for each residue of the antigen. With this approach we were able to compare the overall binding behavior with the behavior of each residue, and then propose antigen residues for favorable mutation. We used a modified $dV$ metric[58], $mdV$, in order to measure the disagreement between the reference clustering (with the whole contact matrix) and each individual residue clustering. Given two clusterings $C1$ and $C2$ of a set $V$ with objects $u$ and $v$ in $V$, $dV$ is computed as follows:

$$dV(C1, C2) = \frac{\sum_{(u,v) \in V \times V} d_{u,v}(C1, C2)}{|V| \cdot (|V| - 1)/2}$$

Where $d_{u,v}(C1, C2)$ is 1 if clusterings $C1$ and $C2$ disagree: $C1(u) = C1(v)\ and\ C2(u) \neq C2(v)$, or $C1(u) \neq C1(v)\ and\ C2(u) = C2(v)$, and 0 otherwise. The modified $dV$ metric, $mdV$, is found by subtracting $dV$ by 1: 1-$dV$ and then normalizing the mean to 0. This results in the $mdV$ indicating less binding contribution if negative and more stable binding contribution if positive, making it more intuitive to visually compare in plots.

In order to have reliable results, we repeated the whole pipeline with different CVAE hyper-parameters. Specifically we used a grid-parameterization varying the number of latent dimensions: d∈(3, 5, 10) and the time delta for the contact tensor definition: delta∈(10, 25, 50) time steps. The CVAE structures we used were fixed, having specular encoder and decoder networks, with 4 layers each with 32 convolutional filters of size 1x7. We trained a total of 9 different CVAE architectures and averaged the clustering results given by each one. We parameterized HDBSCAN clustering algorithm with a minimum cluster size of 50 and both leaf size and minimum samples of 3.

**Free Energy Perturbation.** Based on CVAE results, several systems were selected for mutational binding affinity calculations: insulin DQ8 register3, x-idiotype DQ8 register3, and healthy control DQ8 register 3. Binding affinity changes were computed using Free Energy Perturbation calculations based on previous work[51, 59-62] employing a custom short-range potential on NAMD2. Initial structures for FEP calculations were taken from the last frame of MD simulations based on RMSD analysis showing structural stability (<0.35 Å) for 300+ ns. All FEP calculations were conducted over a course of 34 $\lambda$ windows with at least 200 ps per window for a total of 6.8+ ns per calculation. All FEP calculations were computed with 5 replicas and averaged for the final ΔΔG value. All error bars in this work are 95% CI unless otherwise noted.

All structures are visualized with VMD while all plots are constructed with matplotlib. The SVM used to classify anchor residues was built using the scikit-learn library with a radial basis function kernel, regularization parameter C=30, and kernel coefficient gamma=0.03. Optimal parameters were determined using a grid search.

## Acknowledgements

The authors wish to acknowledge Serena H. Chen, Jeff K. Weber, Sangyun Lee, and Seung-gu Kang for insightful discussions of this work.

## Author Contributions

D.R.B., G.D., L.Z., and G.C. designed the research; D.R.B., G.D., and L.Z. performed research; D.R.B., G.D., J.Y., L.Z., and G.C. analyzed data; D.R.B., G.D., J.Y., R.Z., L.Z., and G.C. conceived the research idea; and D.R.B., G.D., J.Y., L.Z., and G.C. wrote the paper.

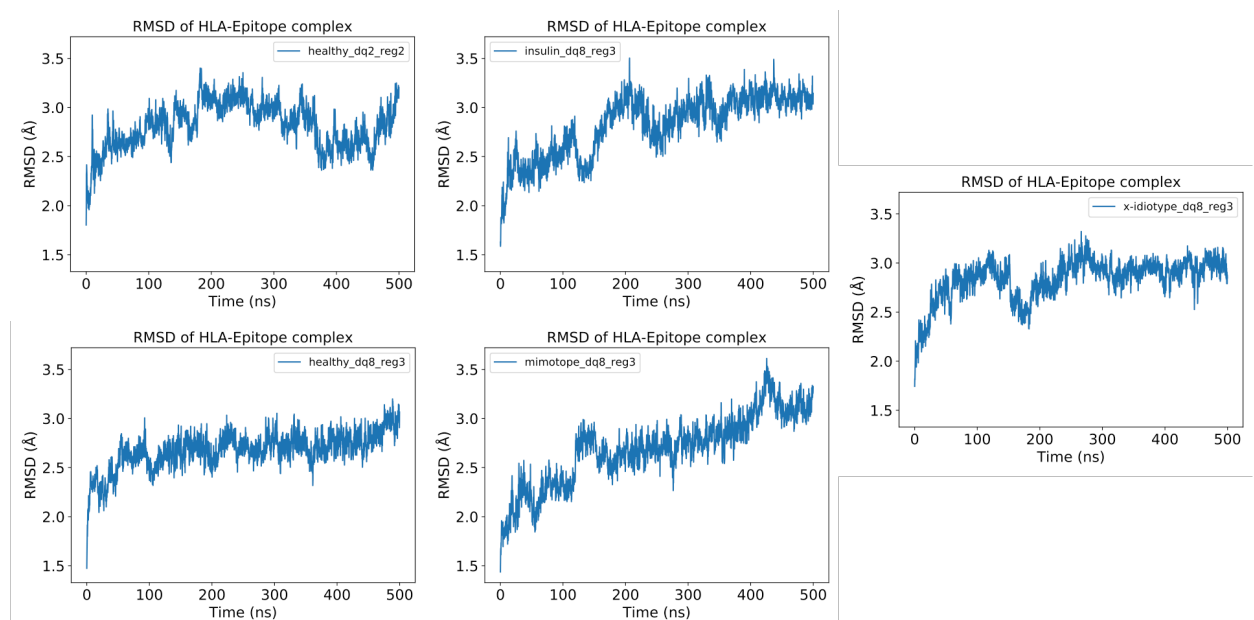

**Figure S1.** Representative RMSD profiles of HLA-antigen systems studied via Free Energy Perturbation. Note: only the HLA-α1 and HLA-β1 subunits are included in the RMSD calculation. The final structure from each trajectory was extracted for FEP calculation.

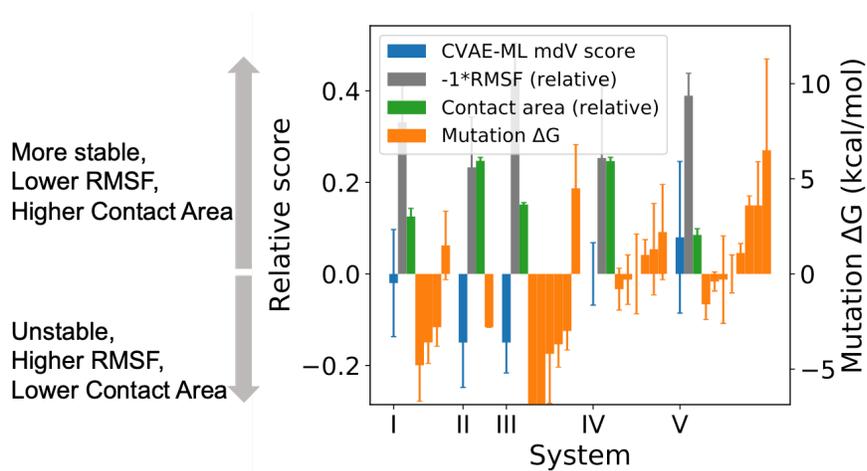

**Fig. S2.** RMSF and relative contact area as predictors of residue binding contributions. As seen in the bar plot, RMSF and contact area values show greater stability for all 5 selected residue systems. However, the systems II and III are predicted not contributing to binding stability and can thus be optimized for more favorable binding affinity. This illustrates that RMSF and relative contact areas do not capture residue contributions to binding stability, while the *mdV* metric used in this work is able to capture contributions to binding stability. All error bars are 95% CI. Residue *mdV*, RMSF, and contact area values are all relative to the 0-mean residue value for the antigen.

**Table S1.** Anchor Residue Determination Test Set.

| PDB ID | HLA Serotype | Antigen Sequence | Expt. Accuracy (Å) |
|---|---|---|---|
| 1fv1 | HLA-DR2a | NPVVHFFKNIVTPRTPPPSQ | 1.9 |
| 1j8h | HLA-DR4 | PKYVKQNTLKLAT | 2.4 |
| 1sjh | HLA-DR1 | PEVIPMFSALSEG | 2.25 |
| 1uvq | HLA-DQ6 | MNLPSTKVSWAAVGGGGSLV | 1.8 |
| 2fse | HLA-DR1 | AGFKGEQGPKGEPG | 3.1 |
| 2nna | HLA-DQ8 | SGEGSFQPSQENP | 2.1 |
| 2q6w | HLA-DR52a | AWRSDEALPLG | 2.25 |
| 4i5b | HLA-DR1 | VVKQNCLKLATK | 2.12 |
| 4is6 | HLA-DR4 | RQLYPEWTEAQRL | 2.5 |
| 4ozh | HLA-DQ2 | APQPELPYPQPGS | 2.8 |
| 4ozi | HLA-DQ2 | QPFPQPELPYP | 3.2 |
| 5ksa | HLA-DQ8 | QPQQSFPEQEA | 2 |
| 5ksu | HLA-DQ2.5 | PVSKMRMATPLLMQ | 2.73 |
| 5ksv | HLA-DQ2.5 | MATPLLMQALPM | 2.19 |
| 5lax | HLA-DRB4 | SKGLFRAAVPSGAS | 2.6 |
| 6biy | HLA-DRB1 | DIFERIASEASRL | 2.05 |
| 6cpl | HLA-DR11 | VDRFYKTLRAEQASQE | 2.45 |
| 6dig | HLA-DQ6 | AGNHAAGILTLGK | 2 |
| 6hby | HLA-DR1 | ARRPPLAELAALNLSGSRL | 1.95 |
| 6u3m | HLA-DQ2.5 | QPMPMPELPYP | 1.9 |

**Table S2. X-idiotype DQ8 Reg3 CVAE data**

| X-idiotype Residue | CVAE count | dv score | dv std dev. | 1-dv | 1-dv relative to mean | mdv:1-dv rel mean-1 | 95% CI |
|---|---|---|---|---|---|---|---|
| 1_CYS | 18 | 0.409 | 0.131 | 0.591 | 1.147 | 0.147 | 0.117 |
| 2_ALA | 18 | 0.371 | 0.078 | 0.629 | 1.221 | 0.221 | 0.070 |
| 3_ARG | 18 | 0.483 | 0.049 | 0.517 | 1.004 | 0.004 | 0.044 |
| 4_GLN | 18 | 0.437 | 0.107 | 0.563 | 1.093 | 0.093 | 0.096 |
| 5_GLU | 18 | 0.493 | 0.08 | 0.507 | 0.984 | -0.016 | 0.072 |
| 6_ASP | 18 | 0.519 | 0.066 | 0.481 | 0.934 | -0.066 | 0.059 |
| 7_THR | 18 | 0.534 | 0.069 | 0.466 | 0.905 | -0.095 | 0.062 |
| 8_ALA | 18 | 0.517 | 0.062 | 0.483 | 0.938 | -0.062 | 0.056 |
| 9_MET | 18 | 0.53 | 0.055 | 0.47 | 0.912 | -0.088 | 0.049 |
| 10_VAL | 18 | 0.509 | 0.066 | 0.491 | 0.953 | -0.047 | 0.059 |
| 11_TYR | 18 | 0.562 | 0.074 | 0.438 | 0.850 | -0.150 | 0.066 |
| 12_TYR | 18 | 0.485 | 0.076 | 0.515 | 1.000 | 0.000 | 0.068 |
| 13_PHE | 18 | 0.505 | 0.068 | 0.495 | 0.961 | -0.039 | 0.061 |
| 14_ASP | 18 | 0.496 | 0.113 | 0.504 | 0.978 | -0.022 | 0.101 |
| 15_TYR | 18 | 0.421 | 0.16 | 0.579 | 1.124 | 0.124 | 0.143 |
| 16_TRP | 18 | 0.487 | 0.128 | 0.513 | 0.996 | -0.004 | 0.115 |

**Table S3. Insulin DQ8 Reg3 CVAE data**

| Insulin Residue | CVAE count | dv score | dv std.dev | 1-dv | 1-dv relative to mean | mdv:1-dv rel mean-1 | 95% CI |
|---|---|---|---|---|---|---|---|
| 1_SER | 18 | 0.515 | 0.06 | 0.485 | 0.839 | -0.161 | 0.048 |
| 2_HSE | 18 | 0.483 | 0.069 | 0.517 | 0.894 | -0.106 | 0.055 |
| 3_LEU | 18 | 0.5 | 0.095 | 0.5 | 0.865 | -0.135 | 0.076 |
| 4_VAL | 18 | 0.266 | 0.131 | 0.734 | 1.270 | 0.270 | 0.105 |
| 5_GLU | 18 | 0.357 | 0.108 | 0.643 | 1.112 | 0.112 | 0.086 |
| 6_ALA | 18 | 0.436 | 0.146 | 0.564 | 0.976 | -0.024 | 0.117 |
| 7_LEU | 18 | 0.336 | 0.135 | 0.664 | 1.149 | 0.149 | 0.108 |
| 8_TYR | 18 | 0.305 | 0.141 | 0.695 | 1.202 | 0.202 | 0.113 |
| 9_LEU | 18 | 0.324 | 0.125 | 0.676 | 1.170 | 0.170 | 0.100 |
| 10_VAL | 18 | 0.41 | 0.084 | 0.59 | 1.021 | 0.021 | 0.067 |
| 11_CYS | 18 | 0.484 | 0.074 | 0.516 | 0.893 | -0.107 | 0.059 |
| 12_GLY | 18 | 0.478 | 0.118 | 0.522 | 0.903 | -0.097 | 0.094 |
| 13_GLU | 18 | 0.482 | 0.1 | 0.518 | 0.896 | -0.104 | 0.080 |
| 14_ARG | 18 | 0.509 | 0.123 | 0.491 | 0.849 | -0.151 | 0.098 |
| 15_GLY | 18 | 0.445 | 0.084 | 0.555 | 0.960 | -0.040 | 0.067 |

**Table S4. Healthy Control DQ8 Reg3 CVAE data**

| Healthy Control Residue | CVAE count | dv score | dv std dev | 1-dv | 1-dv relative to mean | mdv:1-dv rel mean - 1 | 95% CI |
|---|---|---|---|---|---|---|---|
| 1_CYS | 18 | 0.458 | 0.153 | 0.542 | 0.917 | -0.083 | 0.120 |
| 2_ALA | 18 | 0.368 | 0.172 | 0.632 | 1.069 | 0.069 | 0.134 |
| 3_ARG | 18 | 0.427 | 0.172 | 0.573 | 0.970 | -0.030 | 0.134 |
| 4_GLN | 18 | 0.417 | 0.187 | 0.583 | 0.987 | -0.013 | 0.146 |
| 5_ARG | 18 | 0.453 | 0.217 | 0.547 | 0.926 | -0.074 | 0.170 |
| 6_PHE | 18 | 0.348 | 0.197 | 0.652 | 1.103 | 0.103 | 0.154 |
| 7_TRP | 18 | 0.401 | 0.21 | 0.599 | 1.014 | 0.014 | 0.164 |
| 8_SER | 18 | 0.36 | 0.212 | 0.64 | 1.083 | 0.083 | 0.166 |
| 9_GLY | 18 | 0.429 | 0.18 | 0.571 | 0.966 | -0.034 | 0.141 |
| 10_PRO | 18 | 0.403 | 0.218 | 0.597 | 1.010 | 0.010 | 0.170 |
| 11_LEU | 18 | 0.422 | 0.194 | 0.578 | 0.978 | -0.022 | 0.152 |
| 12_PHE | 18 | 0.384 | 0.183 | 0.616 | 1.042 | 0.042 | 0.143 |
| 13_ASP | 18 | 0.416 | 0.2 | 0.584 | 0.988 | -0.012 | 0.156 |
| 14_TYR | 18 | 0.429 | 0.196 | 0.571 | 0.966 | -0.034 | 0.153 |
| 15_TRP | 18 | 0.421 | 0.199 | 0.579 | 0.980 | -0.020 | 0.156 |